\documentstyle{elsart}

\begin{document}

\begin{frontmatter}
\hfill SISSA/99/95/EP

\title{The structure of the ground ring in critical $W_3$ gravity}
\author{Chuan-Jie Zhu}
\address{International School for Advanced Studies, Via Beirut 2--4,
I-34014 Trieste, Italy}
\address{Physics Department, Graduate School,
Chinese Academy of Sciences, P. O. Box 3908,
Beijing 100039, P. R. China}

\begin{abstract}
By explicit calculation, I determine the structure of the ground ring
of the critical $W_3$ gravity and show that there is an $su(3)$
invariant quadratic relation among the six basic elements. By using
this result, I also construct some discrete physical states of the
critical $W_3$ gravity.
\end{abstract}

\end{frontmatter}

The study of $W$ gravity is surely an interesting extension of the
usual two dimensional gravity coupled with matter.  The matter part is
the minimal $W$-matter. Particularly interesting is the borderline
case where the matter part has an integral central charge which is
equal to the rank of the associated Lie algebra. (I will call this case
the critical $W$ gravity.) The study of the simplest case
\cite{lian,bmpa,wittena,wittenb,polya,zhua}, the $D=2$ string theory,
has revealed a lot of interesting structures: the existence of a
ground ring \cite{wittena} and the infinite dimensional Lie algebra
generated by ghost number 0 currents \cite{wittena,wittenb,polya}.
I will report in this letter some results on the extension of these
structures to the critical $W_3$ gravity.

Several papers \cite{popeb,mitb,bmpb,sadov,bmpc,bmpd} have studied the
cohomology problem of $W$ gravity. Nevertheless the results are still
not complete and explicit. The methods used often turn out to be quite
transendential, or by using computer to do most of the calculations,
the results are quite messy and are of little use.  Especially
important is what is the analogous ground ring structure and what is
the infinite dimensional Lie algebra in these $W$ gravity
theories. Many conjectures exist in the literature and very few proofs
and explicit results are offered.  Here I will study the structure of
the ground ring in critical $W_3$ gravity by explicit
calculation. Although most of the calculations are also done by
computer \cite{thie,math}, I have try my best to give the results in
their simplest analytic forms. I will show that the ground ring in
critical $W_3$ gravity is not a free polynomial ring of 6
elements. This ring is actually a polynomial ring of 6 elements module
one quadratic relation. This quadratic relation is $su(3)$
invariant. By exploiting the property of this ground ring, I also
construct some discrete physical states. Partial results have also
been obtained for the algebra generated by the discrete states and the
generalization to other critical $W$ gravities, but I will not report
these results here.

As usual I deal with only one chiral sector and restrict my attention
to the critical case. I will speak only the prime physical states or
the relative physical states in a generalized relative cohomology.
For the pure $W_3$ gravity, the complete physical states have been
obtained in \cite{popeb} but there is no interesting symmetry
structures, just as in the case with the usual pure gravity
(i.e. $W_2$ gravity).  For the $W_3$ gravity coupled with (the $W_3$)
minimal matter, the study and enumeration of all the physical states
is complicated by the problem of decoupling all the null matter
states, see \cite{bmpd} and refernces therein.  The critical $W_3$
gravity is easier and is also more interesting. This theory has been
studied in \cite{mitb,bmpc}.

The basic fields of the critical $W_3$ gravity are the two free matter
fields $X_1(z)$ and $X_2(z)$, two Liouville fields $\phi_1(z)$
and $\phi_2(z)$, two pairs of ghost anti-ghost fields $\big( b(z),
c(z) \big)$ and $\big(\beta(z), \gamma(z) \big)$ associated with
the spin 2 and 3 generators of the $W_3$ algebra.  From these fields
we can construct the following stress-energy tensors and two
spin-3 generators:
\begin{eqnarray}
 T_X & = & - {1\over 2} \big(\partial_z X_1(z)\big)^2 -
- {1\over 2} \big(\partial_z X_2(z)\big)^2, \\
T_{\phi} & = & - {1\over 2} \big(\partial_z \phi_1(z)\big)^2 -
- {1\over 2} \big(\partial_z \phi_2(z)\big)^2 + \sqrt{2}
\partial_z^2\phi_1(z) + \sqrt{6} \partial_z^2\phi_2(z), \\
T_{bc} & = &\quad 2\, \partial_zc(z)\, b(z) + c(z)\, \partial_z b(z),
\label{deftt} \\
T_{\beta\gamma} & = &\quad 3\, \partial_z\gamma(z) \, \beta(z) +
2 \gamma(z)\, \partial_z\beta(z) ,
\end{eqnarray}
and
\begin{eqnarray}
W_X & = &\quad  { i\over 6 } \Big( 3\, \big(\partial_z X_1(z)\big)^2 -
\big( \partial_z X_2(z)\big)^2 \Big)\, \partial_z X_2(z) , \\
W_{\phi} & = &\quad  { i \over 24} \Big(
3\, \big( \partial_z\phi_1(z) \big)^2\, \partial_z\phi_2(z) -
\big( \partial_z \phi_2(z)\big)^3  +
3\sqrt{6}\partial_z\phi_2(z)\,\partial_z^2\phi_2(z) \label{defww}
\nonumber \\
& - & 3\big(\sqrt{6}\, \partial_z\phi_1(z) +
2\sqrt{2}\, \partial_z \phi_2(z)\big) \partial_z^2\phi_1(z)
+ 6\sqrt{3}\, \partial^3_z\phi_1(z) - 6\,\partial_z^2\phi_2(z) \Big) .
\nonumber
\\ & &
\end{eqnarray}
The two pairs $(T_X,W_X)$ and $(T_{\phi}, W_{\phi})$ satisfy the
nonlinear $W_3$ algebra \cite{zam} with central charge $c=2$ and 98
respectively.  The combined matter-Liouville stress-energy tensor
$T_{X\phi}=$ $T_X + T_{\phi}$ then has central charge 100 which is the
critical central charge of the $W_3$ algebra \cite{min}.  It is then
possible to have a nilpotent BRST for this
matter-Liouville-ghost-antighost system.  Actually two different
nilpotent BRST charges are known \cite{min,mita,shen}.  The one
obtained in \cite{mita,shen} is better suited for the study of the
critical $W_3$ gravity. This is because the standard physical
tachyon-like states have a simple form with respect to this BRST
charge. We will use the following BRST charge for all our explicit
calculation, although the results obtained don't depend on the exact
form of the BRST charge.  The BRST charge $Q$ is the contour
integration of the BRST current $j(z)$: $Q\equiv \oint_0 [\d z] j(z)$
and
\begin{eqnarray}
 j(z) & = & c(z)\, \Big( T_X(z) + T_{\phi}(z) + \partial_z c(z)\, b(z)
+ T_{\beta\gamma}(z) \big) \nonumber \\ & + & \sqrt{6}\,
\gamma(z)\Big( W_X(z) + 4i\, W_{\phi}(z) \Big) + 3\, \Big( T_X(z) -
T_{\phi}(z)\Big)\, b(z)\,\gamma(z)\, \partial_z\gamma(z) \nonumber \\
& - & {3\over 2} \Big(2\,\gamma(z)\, \partial_z^3\gamma(z) -
3\,\partial_z \gamma(z)\, \partial_z^2\gamma(z) \Big)\, b(z) .
\label{eq:currentz}
\end{eqnarray}
Here we have rescaled $\gamma$ and $\beta$ by a factor $\sqrt{3}$ to
avoid some unnecessary $\sqrt{3}$ factors in the expression of ground
ring states (see below), comparing with \cite{mitb}.

Denoting the six different Weyl transformations of the Weyl group of
the $su(3)$ Lie algebra as $\sigma_i$, $ i = 0,\cdots,5$, the six
tachyon-like physical states are given as follows
\begin{equation}
 V_{p_X,p_{\phi}}(z) = c(z)\,\gamma(z)\,\partial_z\gamma(z)\,
e^{ i \, \sigma_i\,(p_{\phi} - 2\, \rho )\cdot X(z) +
p_{\phi}\cdot \phi(z)} , \qquad
i = 0,\cdots,5 . \label{solutiona}
\end{equation}
For generic momentum $p_{\phi}$, the above tachyon-like states are the
only physical states with two arbitrary parameters. Later we will
identify some one parameter continuous physical states.

Ground ring was introduced in the study of $D=2$ string theory by
Witten \cite{wittena}. However this concept has more general usage and
can be extended to the study of other theories. The ground ring for
$W_3$ gravity has been discussed in \cite{mitb} and explicit formulas
are also given for the basic elements. However, little was known about
the propertites of the ring structure and their applications little
exploited.  Let us first recall some basics about the ground ring in
$D=2$ string theory.  In $D=2$ string theory, there exist two basic
physical states (with ghost number 0) which form the two dimensional
multiplet, the lowest nontrivial representation of the $su(2)$
algebra. Explicitly these two states are given as follows:
\begin{eqnarray}
x & = & (c\, b + S_1(i X^-) )\, e^{i\,  X^+}, \\
y & = & (c\, b + S_1(-i X^+) )\, e^{-i\,  X^+} ,
\end{eqnarray}
where $X^{\pm} = {1\over \sqrt{2}}\,(X \pm i\, \phi)$ and $S_1$ is the
first Schur polynomial\footnote{Explicitly we have:
$$ S_k(i\delta\cdot X(w)) e^{i\delta \cdot X(w)} = { 1\over k!} \,
\partial_w^k (e^{i\delta \cdot X(w)}). $$}.  The ground ring is just
the {\bf free} polynomial ring in $x$ and $y$ if we define the
multiplication as the usual normal ordering at the same point module
BRST exact terms.  The states with a fixed momentum $p_{\phi}$ form a
$2\,j+1$ dimensional representation of the $su(2)$ algebra and these
states corresponds to the ring elements with a fixed degree $2\,j+1$,
where $j$ is an integer or half integer. In critical $W_3$ gravity, we
need two basic representations of the $su(3)$ algebra which are
denoted as ${\bf 3}$ (with highest weight $\lambda_1$) and $\bar{{\bf
3}}$ (with highest weight $\lambda_2$) or $\{1,0\}$ and $\{0,1\}$ in
the notation of Dykin index.  We also denote the elements of ${\bf 3}$
as $x_i$, $i=1,2,3$ and those of $\bar{{\bf 3}}$ as $y_i$,
$i=1,2,3$. These states have the following general form:
\begin{eqnarray}
x_i & = &  \big( \hbox{something of ghost number $0$} \big) \times
 e^{ i \, m_i\cdot X - \lambda_2 \cdot \phi } , \\
y_i & = &  \big( \hbox{something of ghost number $0$} \big) \times
 e^{- i \, m_i\cdot X - \lambda_1 \cdot \phi } ,
\end{eqnarray}
where $m_i$'s are the three weights of the representation ${\bf 3}$
(with highest weight $\lambda_1$). Here we have used the fact that the
weights of $\bar{{\bf 3}}$ can be obtained from the weights of {\bf 3}
by multiplying $-1$. We will also denote $x_i$ as
$O^{\lambda_1}_{m_i}$ and $y_i$ as $O^{\lambda_2}_{-m_i}$ where the
superscript denotes the representation and the subscript denotes the
$X$-momentum.  The explicit expression of these states can be obtained
by using computer.  We have take a little time to write the resulting
expression in the following form:
\begin{eqnarray}
 x_1 & \equiv & O^{\lambda_1}_{\lambda_1} = \Big( \big(
 c\,b\,\gamma\,\beta + c\, b' + {3\over 2}\,\gamma\,\beta + {7\over
 2}\,\gamma'\,\beta - 7\, b\,b'\,\gamma\,\gamma' \nonumber \\ & - &
 {5\over 2}\,b\,\gamma\,\gamma'\,\beta - {21\over 4}\,b\,\gamma'' -
 {1\over 2}\, c\,\beta + {1\over 2}\,c\,b\,b'\,\gamma - 2
 \,b'\,\gamma' -{3\over 4}\,b''\,\gamma \big) \nonumber \\ & + &
 {3\over 2} \, (c\,b + \gamma\, \beta + 3\, b\,\gamma' + b'\,\gamma)
 \, S_1(i\lambda_1\cdot X) + {3\over 2} (\gamma\,\beta - 2\,b
 \,\gamma') \, \lambda_2\cdot \phi' \nonumber \\ & + & 3\, (1 +
 b\,\gamma)\big( S_2(i\lambda_1\cdot X) + {1\over 2}\,
 S_1(i\lambda_1\cdot X)\, \lambda_2\cdot \phi' \big) \nonumber \\ & +
 & 3\, (1 - 2\, b\,\gamma)( - {1\over 2}\, (\lambda_1 -
 \lambda_2)\cdot \phi'\, \lambda_1\cdot \phi' + \lambda_1\cdot \phi''
 ) \Big) \, e^{ i\lambda_1 \cdot X - \lambda_2 \cdot \phi} ,
\label{eq:xxxa}
\end{eqnarray}
where $S_i$'s are Schur polynomials. The two other states in {\bf 3}
can be obtained by $su(3)$ action. The resulting expression is the
same as above but with $S_i(i\lambda_1\cdot X)$ change to
$S_i(i\, m_i\cdot X)$ in accordance with
the exponetial factor.  The expression for the states
$y_i$ in $\bar{\bf 3}$ can also be obtained from $(\ref{eq:xxxa})$
by noting the following transformation of the BRST charge:
\begin{equation}
\gamma  \to - \gamma, \qquad  \beta  \to - \beta, \qquad
 X  \to - X, \qquad
\lambda_1\cdot \phi  \leftrightarrow \lambda_2 \cdot \phi.
\label{symm}
\end{equation}
This transformation transforms one nontrivial physical states into
another nontrivial physical states. Explicitly we have:
\begin{eqnarray}
y_1 & \equiv & O^{\lambda_2}_{-\lambda_1} = \Big( \big(
 c\,b\,\gamma\,\beta + c\, b' + {3\over 2}\,\gamma\,\beta + {7\over
 2}\,\gamma'\,\beta - 7\, b\,b'\,\gamma\,\gamma' \nonumber \\ & + &
 {5\over 2}\,b\,\gamma\,\gamma'\,\beta + {21\over 4}\,b\,\gamma'' +
 {1\over 2}\, c\,\beta - {1\over 2}\,c\,b\,b'\,\gamma + 2
 \,b'\,\gamma' +{3\over 4}\,b''\,\gamma \big) \nonumber \\ & + &
 {3\over 2} \, (c\,b + \gamma\, \beta - 3\, b\,\gamma' - b'\,\gamma)
 \, S_1(-i\lambda_1\cdot X) + {3\over 2} (\gamma\,\beta + 2\,b
 \,\gamma') \, \lambda_1\cdot \phi' \nonumber \\ & +& 3\, (1 -
 b\,\gamma)( S_2(-i\lambda_1\cdot X) + {1\over 2}\,
 S_1(-i\lambda_1\cdot X)\, \lambda_1\cdot \phi' ) \nonumber \\ & +&
 3\, (1 + 2\, b\,\gamma)( {1\over 2}\, (\lambda_1 - \lambda_2)\cdot
 \phi'\, \lambda_2\cdot \phi' + \lambda_2\cdot \phi'' ) \Big) \, e^{ -
 i\lambda_1 \cdot X - \lambda_1 \cdot \phi} .
\label{eq:xxxb}
\end{eqnarray}
The other two $y_i$'s can be obtained from this one by $su(3)$ action.
This changes only the arguments of the the Schur polynomials and the
exponential factor.

Similar to $D=2$ string theory, the ground ring of the critical $W_3$
gravity is the polynomial ring in six elements $x_i$ and $y_i$,
$i=1,2,3$.  The difference is that this ring is {\bf not free},
i.e. there is a polynomial relation between the six basic elements
$x_i$ and $y_j$:
\begin{equation}
x_1\cdot y_1 + x_2\cdot y_2 + x_3\cdot y_3 =  0. \label{eq:ground}
\end{equation}
This is equivalent to say that the tensoring of {\bf 3} and $\bar{
{\bf 3}}$ gives only {\bf 8}. The state which could give rise to the
trival representation {\bf 1} is proportinal to the left hand side of
eq. (\ref{eq:ground}), but it is vanishing.  It is quite difficult to
prove the above equation directly.  I will talk about the proof in a
moment, but let us first explore the the consequence of the above
equation.  By taking tensor product of the two basic representations
we can obtain any representations. Because of eq. (\ref{eq:ground}),
the structure of tensor product is greatly simplified. We have only
the following rules:
\begin{equation}
\{n_1,n_2\} \otimes \{m_1,m_2\} = \{n_1+m_1,n_2+m_2\},
\label{tensorp}
\end{equation}
i.e. the products of the states in two highest
representations with highest weights $\Lambda_1$ and $\Lambda_2$ give
only states which form another highest weight representation with
highest weight $\Lambda = \Lambda_1 + \Lambda_2$. Presumably this
procedure gives all the physical states with ghost number 0 (more
about this point at the end of this paper). So we
conclude that only for $p_{\phi} = -(n_2 \, \lambda_1 +
n_1\,\lambda_2)$, ($n_1,n_2$: non-negative integers) there exist ghost
number 0 relative physical states.  These states form a highest weight
representation of $su(3)$ with highest weight $\Lambda = (n_1 \,
\lambda_1 + n_2\,\lambda_2)$ and their expression can be obtained from
the highest state $O_{\Lambda}^{\Lambda} \equiv
(O^{\lambda_1}_{\lambda_1})^{n_1}\cdot
(O^{\lambda_2}_{\lambda_2})^{n_2}$ by repeatedly using $su(3)$
actions.

Defining two operators
\begin{eqnarray}
a_+(n,m)   & \equiv & \, [Q, -i\,(n + 2)\,\alpha_1\cdot X
+ m\,\alpha_2\cdot\phi ], \\
a_-(n,m)   & \equiv & \, [Q, -i\,(n + 2)\,\alpha_2\cdot X
+ m\,\alpha_2\cdot\phi ],
\end{eqnarray}
one can prove that the multiplication of $a_+(n_1,m_1)$ and
$a_-(n_1,m_2)$ with $O^{\Lambda}_{\lambda, q(\lambda)}$, having
$\Lambda=$ $n_1\,\lambda_1 + n_2\,\lambda_2$ and $\lambda =$
$m_1\,\lambda_1 + m_2\,\lambda_2$, give rise to new nontrivial
relative states with ghost number 1. Actually these states belong to
the irreducible representation $\Lambda + \rho$ and its highest weight
state is
\begin{eqnarray}
\bar{O}_{\Lambda+ \rho}^{\Lambda + \rho} & \equiv & \,\,
[ E_{\rho}, a_+]\cdot O_{\Lambda}^{\Lambda} \nonumber \\
& =  & \,\,[E_{\rho}, a_+\cdot O_{\Lambda}^{\Lambda}  ].
\end{eqnarray}
This gives some physical states of ghost number 1 with Liouville
momentum $p_{\phi} = -(n_2 \, \lambda_1 + n_1\,\lambda_2)$.  These
states don't exhaust all the discrete phsysical states in the critical
$W_3$ gravity with ghost number 1.\footnote{I would like to thank
K. Pilch and P. Bouwknegt for pointing out this to me.}

The physical states with ghost number 2 can be similarly constructed
and we have the following results: the physical states (having ghost
number 2 and fixed Liouville momentum $p_{\phi} = -(n_2 \, \lambda_1 +
n_1\,\lambda_2)$ form two irreducible representations $\Lambda + 3\,
\lambda_1 $ and $\Lambda + 3\, \lambda_2$ which are generated by the
following two highest weight states:
\begin{eqnarray}
\underline{Y}_{\Lambda + 3 \, \lambda_1}^{\Lambda+3\,\lambda_1}
& \equiv &\,\,
[E_{\alpha_1}, [ E_{\rho}, a_+] ]\cdot O_{\Lambda}^{\Lambda}
\nonumber \\
& =  &\,\, [E_{\alpha_1}, [E_{\rho},
a_+\cdot  O_{\Lambda}^{\Lambda}  ] ], \\
 \underline{Y}_{\Lambda+3 \, \lambda_2}^{\Lambda+3 \, \lambda_2}
& \equiv & \,\,
[E_{\alpha_2}, [ E_{\rho}, a_+] ]\cdot  O_{\Lambda}^{\Lambda}
\nonumber \\
& =  &\,\, [E_{\alpha_2}, [E_{\rho},
a_+\cdot  O_{\Lambda}^{\Lambda}  ] ] .
\end{eqnarray}
Actually the physical states at the boundary of the above
representations can also be promoted to one parameter continuous
physical states, just like the boundary states (with ghost number 1)
in $D=2$ string theory are actually tachyon states. Here in critical
$W_3$ gravity, the ghost number of continuous states can be ranged
from 2 to 4. (If we take into account also the absolute states, the
range is from 2 to 6.)

Finally let us also mention that the discrete states with ghost number
3 are in representation $\Lambda+ 2\, \rho$ and can be generated by
the following highest weight state
\begin{eqnarray}
Y_{\Lambda + 2\,\rho}^{\Lambda +2\, \rho} & \equiv &\,\,
[E_{\alpha_2}, [E_{\alpha_1}, [ E_{\rho}, a_+]] ]\cdot
O_{\Lambda}^{\Lambda}
\nonumber \\
& =  &[E_{\alpha_2}, [E_{\alpha_1}, [E_{\rho},
a_+\cdot  O_{\Lambda}^{\Lambda}  ]] ] \propto c\,\gamma\,\gamma'\,
 e^{ i\,(\Lambda + 2 \, \rho)\cdot X - (n_2\, \lambda_1 + n_1 \,
\lambda_2)\cdot \phi }  .
\end{eqnarray}
Presumbably there exists no relative discrete states with ghost number
4 or greater and having the above mentioned Liouville momentum.

In the above, we have studied only the physical states having the same
Liouville momentum as the ground ring elements. There are about five
copies of these states having other Liouville momentum and ghost
number, just like there are minus states in addition to plus states in
$D=2$ string theory. A complete classificaton of all the physical
states for the critical $W_3$ gravity is not available. Now let me
return to the proof of eq. $(\ref{eq:ground})$.

In order to prove eq. $(\ref{eq:ground})$, we use the following
theorem:
\begin{itemize}
\item If the current $V^{(1)}(w)$ of a BRST invariant state $V(w)$ is
zero module BRST exact and total derivative terms, i.e. $V^{(1)}(w)$
can be written as follows:
\begin{equation}
V^{(1)}(w) = \{  Q, \eta^{(1)} ] + \partial_w\, \eta^{(2)}(w) ,
\label{eq:cccp}
\end{equation}
then we have
\begin{equation}
\partial_w\, \Big( V(w) - \{ Q, \eta^{(2)}(w) ] \Big) = 0,
\end{equation}
i.e. $V(w)$ is either a BRST exact state
or proportional to the trivial BRST invariant state $1$.
\end{itemize}
This theorem can be proved easily by using contour deformation in
conformal field theory. The only requirement is that the BRST charge
should be written as a contour integration of a dimension $1$ current
and that the (anti-) commutator of $Q$ with $b(w)$ gives the total
stress energy tensor. This is satisfied by the BRST charge given in
$(\ref{eq:currentz})$.

By using this theorem, one proves eq. $(\ref{eq:ground})$ if one can
show that the corresponding
current of the left hand side is the form of $(\ref{eq:cccp})$.
This is not so difficult
with our explicit expression for $x_i$ and $y_i$ given in
eqs. $(\ref{eq:xxxa})$ and $(\ref{eq:xxxb})$. However the current
\begin{eqnarray}
j_i & \equiv & \oint_w [\d z] \, b(z)\, (x_i\cdot y_i)(w)
\nonumber \\
& = & \oint_w [\d z]
\big(b(z)\,x_i(w)\big) \cdot y_i(w) + x_i(w)\cdot \oint_w [\d z]
b(z)\, y_i(w) ,
\end{eqnarray}
still contains something over 100 terms. After
subtracting all the total derivative and BRST exact terms, I found
that there are only two independent $j_i$, i.e. there is a linear
relation among the three $j_i$'s:
\begin{equation}
j_1 + j_2 + j_3 = 0. \label{cuid}
\end{equation}
This proves
(\ref{eq:ground}). It may be helpful to display some terms of $j_i$'s
here to see how eq. $(\ref{cuid})$ is satisfied. We have
\begin{eqnarray}
j_1 & = & \Big( \, {9\over 2}\,\partial^2\tilde{X}_1\,
\rho\cdot \partial \phi\, b\,\gamma\,\beta
+ \big(3 \,\partial \tilde{X}_1 \, \rho\cdot \partial \phi -
{15\over 2}\, \partial^2 \tilde{X}_1 \big) \, b\,\gamma\, \beta'
\nonumber \\
& + & {27 \over 4} \, \partial \tilde{X}_1\,
\rho\cdot \partial \phi\, b\,\gamma'\,\beta  -
 {45\over 4} \big(2\,b\,\gamma'\,\beta'
+ 2\,b\,\gamma\,\beta''+ b\,\gamma''\,\beta\big) \partial \tilde{X}_1
\nonumber \\
& - &  \Big( 9 \,(\lambda_1 - \lambda_2)\cdot \partial \phi \,
\partial \tilde{X}_1 + {15\over 4}\, (\partial \tilde{X}_1)^2
 - {15\over 2} \partial \tilde{X}_1\, \partial \tilde{X}_2
\nonumber \\
& - & {15\over 2}\, (\partial \tilde{X}_2)^2 \big)\,b\,b'\,\gamma'
+ (\cdots)\,b\,b'\,\gamma +
(\cdots)\, \beta + (\cdots) \, b \Big)\, e^{ - i\rho\cdot \phi} ,
\end{eqnarray}
where $\tilde{X}_1 = i\,\lambda_1\cdot X$ and $\tilde{X}_2 = -
i\,\lambda_2 \cdot X$. Notice that the above expression changes to
$-j_1$ under the transformation $(\ref{symm})$.  This is understandable
because $x_1 \cdot y_1$ is defined up to a sign and $x_1 \cdot y_1
=\pm y_1 \cdot x_1 $. The other two currents $j_2$ and $j_3$ can be
obtained from the above expression by the following substitutions:
\begin{equation}
\tilde{X}_1 \to \tilde{X}_2, \qquad \qquad
\tilde{X}_2 \to - \tilde{X}_1 - \tilde{X}_2,
\end{equation}
or
\begin{equation}
\tilde{X}_1 \to - \tilde{X}_1 - \tilde{X}_2, \qquad \qquad
\tilde{X}_2 \to \tilde{X}_1 .
\end{equation}
The terms linear in $\tilde{X}_1$ or
$\tilde{X}_2$ cancell outomatically in eq. $(\ref{cuid})$. We have also
proved that other terms cancell with each other.

Actually one can argue that there must exist some relations such as
eq. $(\ref{eq:ground})$ from other sources \cite{bmpc}.  One way to
prove this relation
is to have some similar existence and unique theorem about the
spectrum of physical states in $W_3$ gravity. In \cite{bmpc}, some
theorems and conjectures have been given.  It's quite difficult to
understand their arguments (at least for me).
Our explicit calculation actually solved
the cohomology problem of critical $W_3$ gravity by explicit
construction. We found that by taking tensor product one gets only
once each representation, i.e. for a fixed Liouville momentum, there
is only one irreducible finite-dimensional $su(3)$ ground ring
module. As noted in \cite{bmpc}, there may also exist some other
ground ring elements which cann't be obtained by taking tensor
product of the two basic representations (this may appear in the
interior of the Weyl chamber). To settle this question explicitly,
I have compute all the possible physical states of ghost number 0 and
momentum $p_{\phi}=-(\lambda_1 + \lambda_2)$ by brute force by using
computer. There are only 8 independent non-trivial physical states.
This shows that  there are no other physical states other that those
obtained from tensor product of the two basic representations.
This result has also been prove by K. Pich and P. Bouwknegt by using
a different programme to do the explicit computations.

It's quite difficult to extend our method to other more complicated
critical $W_n$ gravity theories. Nevertheless the results can be easily
generalized to other critical $W$ gravities associated with any simply
laced Lie group.  We have: {\it the ground ring in any critical $W$
gravity is the polynomial ring of all the states of the basic
representations module relations which would give non-maximal
representations when taking tensor product.  In other words, the
ground ring exists and its structure is such that the product of any
two highest weight representations gives only one highest weight
representation.} For a fixed Liouville momentum, there is only one
irreducible finite-dimensional ground ring module. I hope there would
be a different way to arrive at (or to disprove) these conjectures.

\ack{I would like to thank Prof. P. Bouwknegt, Prof. K. Pilch and
Prof. C. Pope for discussions. This work is supported in part by
Pan-Deng-Ji-Hua special project and Chinese Center for Advanced
Science and Technology (CCAST).}

\end{document}